\documentclass[a4paper,UKenglish,cleveref, autoref, thm-restate]{lipics-v2021}

\usepackage{amsthm}
\usepackage{amsmath}
\usepackage{amsfonts}
\usepackage{graphicx}
\usepackage{enumitem}
\usepackage{microtype}

\usepackage{algorithm}
\usepackage{algorithmicx}
\usepackage{algpseudocode}
\usepackage{cite}

\title{Crane Scheduling Problem with Energy Saving} 

\author{Yixiong Gao}{Department of Computer Science, City University of Hong Kong, Kowloon, Hong Kong SAR, China \and \url{https://yixionggao.com/}}{yixiong.gao@my.cityu.edu.hk}{https://orcid.org/0009-0008-8846-3873}{}

\author{Florian Jaehn}{Institute for Management Science and Operations Research, Helmut Schmidt University, Hamburg, Germany \and \url{https://www.hsu-hh.de/or/en/jaehn-en}}{florian.jaehn@hsu-hh.de}{}{}

\author{Minming Li}{Department of Computer Science, City University of Hong Kong, Kowloon, Hong Kong SAR, China \and \url{https://www.cs.cityu.edu.hk/~minmli/}}{minming.li@cityu.edu.hk}{https://orcid.org/0000-0002-7370-6237}{}

\author{Wenhao Ma}{Department of Computer Science, Hangzhou Dianzi University, Hangzhou, Zhejiang, China \and \url{}}{23051127@hdu.edu.cn}{https://orcid.org/0009-0005-3636-2278}{}

\author{Xinbo Zhang}{Department of Decision Analytics and Operations, City University of Hong Kong, Kowloon, Hong Kong SAR, China \and \url{}}{xinbo.zhang@cityu.edu.hk}{https://orcid.org/0000-0002-8786-2708}{}

\authorrunning{Y. Gao et al.}

\Copyright{Yixiong Gao, Florian Jaehn, Minming Li, Wenhao Ma and Xinbo Zhang}



\EventEditors{John Q. Open and Joan R. Access}
\EventNoEds{2}
\EventLongTitle{42nd Conference on Very Important Topics (CVIT 2016)}
\EventShortTitle{CVIT 2016}
\EventAcronym{CVIT}
\EventYear{2016}
\EventDate{December 24--27, 2016}
\EventLocation{Little Whinging, United Kingdom}
\EventLogo{}
\SeriesVolume{42}
\ArticleNo{23}

\begin{document}

\maketitle

\begin{abstract}
During loading and unloading steps, energy is consumed when cranes lift containers, while energy is often wasted when cranes drop containers. By optimizing the scheduling of cranes, it is possible to reduce energy consumption, thereby lowering operational costs and environmental impacts. In this paper, we introduce a single-crane scheduling problem with energy savings, focusing on reusing the energy from containers that have already been lifted and reducing the total energy consumption of the entire scheduling plan. We establish a basic model considering a one-dimensional storage area and provide a systematic complexity analysis of the problem. First, we investigate the connection between our problem and the semi-Eulerization problem and propose an additive approximation algorithm. Then, we present a polynomial-time Dynamic Programming (DP) algorithm for the case of bounded energy buffer and processing lengths. Next, adopting a Hamiltonian perspective, we address the general case with arbitrary energy buffer and processing lengths. We propose an exact DP algorithm and show that the variation of the problem is polynomially solvable when it can be transformed into a path‑cover problem on acyclic interval digraphs. We introduce a paradigm that integrates both the Eulerian and Hamiltonian perspectives, providing a robust framework for addressing the problem.
\end{abstract}

\newpage

\section{Introduction}
\label{sec1:introduction}
\begin{figure}[htbp]
    \centering
    \captionsetup{justification=centering}
	\includegraphics[width=\textwidth]{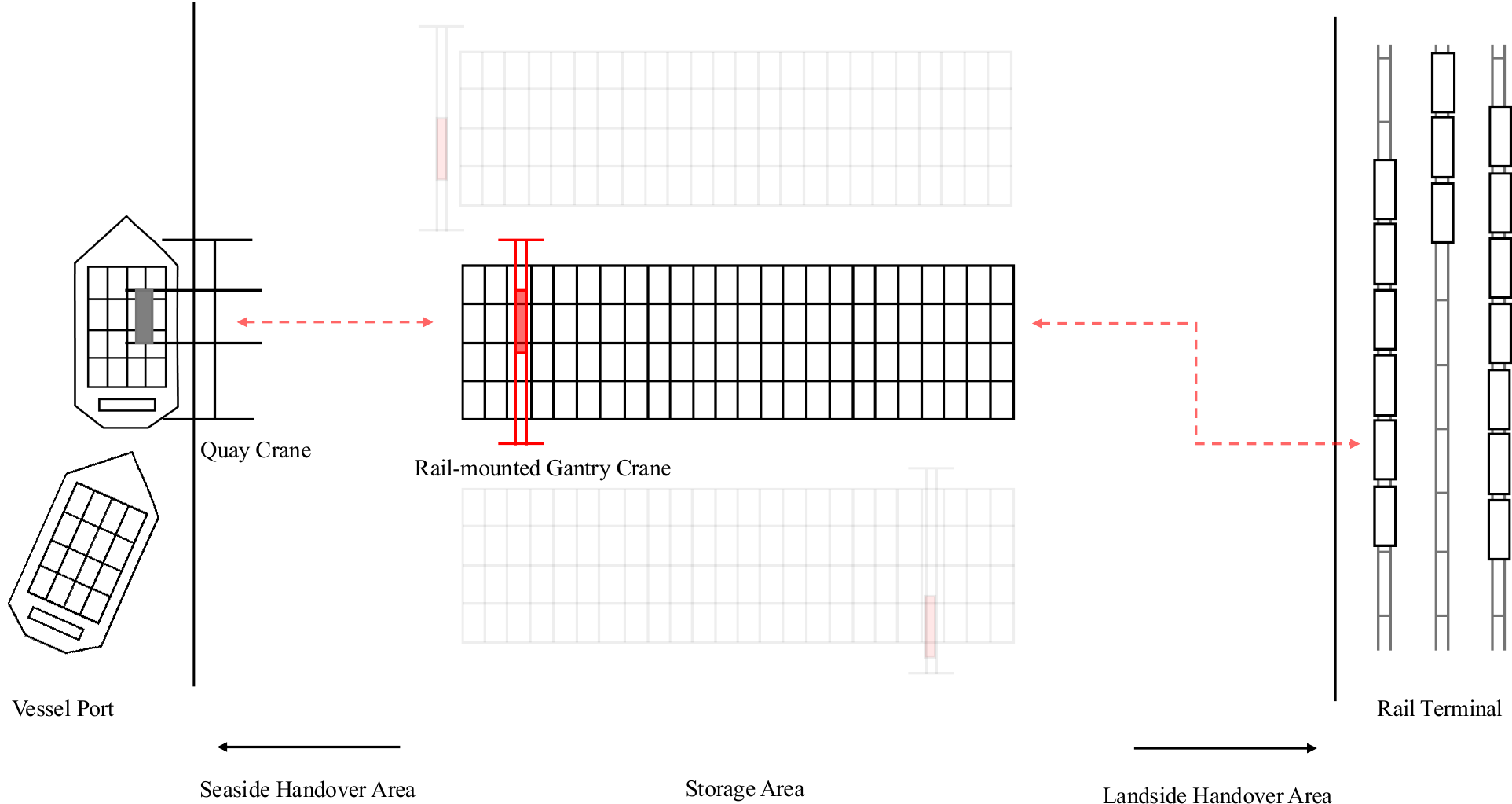}
	\caption{Layout at Container Terminal} 
	\label{layout}
\end{figure}

Port cities have advanced logistics and transportation industries, with their economic development partly relying on high cargo throughput. Managers and experts need to create efficient schedules to store and transport containers in these cities' highly automated container terminals. To handle the multitude of assignments and scheduling problems faced daily, managers must devise strategies that ensure smooth operations \cite{hhla2014}. We refer to the layout of Container Terminal (CT) \cite{ehleiter2016housekeeping} and extract its features for our subsequent scheduling model. Figure \ref{layout} describes the typical container movement process through a CT. Based on \cite{hhla2014}, when a vessel arrives at the port, several quay cranes are busy unloading and loading containers, which are then transported to the storage area. For containers temporarily stored in the storage area, gantry cranes will either transport them to the seaside for transport to other vessels, or to the landside for departure from the rail terminal. In this paper, we focus on the gantry crane scheduling process in the storage area.

On the other hand, with the advocacy of green industrial ideas in recent years, both the public and companies have increasingly focused on low-carbon environmental measures. Advanced logistics industries aim to reduce energy consumption when handling containers. In the case of gantry cranes, Parise et al. mentioned that the largest share of energy is required for lifting a container \cite{parise2014port}. This was further highlighted in HHLA 2019 \cite{HHLA19}. When cranes lower containers, part of this energy is released and often wasted. Therefore, each container's lift and lowering operations consume and release a significant amount of energy. The crane can utilize this energy more efficiently if the energy released during the lowering of a container is stored and used to lift the next container. The problem we analyze here is to potentially build a valid schedule to reduce the amount of energy consumed and, thus, reduce $CO_2$ emissions along with operating costs in the container terminal.

So far, several studies have focused on optimizing the layout at container terminals. Yu et al. \cite{yu2022yard} surveyed yard management in automated container terminals for the last 20 years. For the crane scheduling problem itself, there are currently two main research directions: minimizing the schedule length \cite{ehleiter2018scheduling,oladugba2023new} or minimizing the number of violated restrictions \cite{park2010real,vallada2023models}. Briskorn et al. \cite{briskorn2019generator} developed a generator capable of producing test instances for arbitrary terminal layouts. Moreover, Jaehn et al. \cite{jaehn2020approximation} proposed a twin crane scheduling model, which adapts to real scheduling situations in harbors. Oladugba et al. \cite{oladugba2023new} addressed the twin yard crane scheduling problem in automated container terminals to minimize makespan. They propose a modified Johnson's algorithm for a mixed integer programming model. Vallada et al. \cite{vallada2023models} studied the yard crane scheduling problem. They considered the problem as a pickup routing problem and a one-machine scheduling problem to formulate the problem. Yin et al. \cite{yin2025scheduling} investigated the scheduling of twin automated stacking cranes for job assignment at an automated container terminal, defining dedicated handover zones at either end of each yard block. However, with the increasing awareness of industrial ecology \cite{jaehn2016sustainable}, the concept of green crane scheduling has been proposed \cite{froese2014green, wilmsmeier2016energy, geerlings2018opportunities, weckenborg2024energy}. Despite this, studies on crane scheduling with an energy-saving condition are very rare. The most promising idea is to use the energy from a currently lifted container for subsequent actions. For example, in 2008, Flynn et al. \cite{flynn2008saving} described an energy-saving mechanism called the `flywheel', which has high energy efficiency but quickly loses energy over time. This paper is based on the assumption that, for a short period, most of the stored energy can be used to lift containers; after this period, the stored energy will be released. This is based on the observation that the flywheel reaches its maximum power with a short delay after energy is put into it because it is still spinning up. Thereafter, there is a significant energy loss in a short amount of time. Since most of the energy is lost right after the flywheel reaches its maximum power, we consider the loss of energy to be abrupt, although some energy remains in practice. HHLA \cite{HHLA19} confirmed that lifting containers consumes a certain amount of energy from the gantry crane, while lowering them releases energy---a fact that has been recognized in earlier studies as well \cite{parise2014port}. Combining these two aspects can significantly save energy during scheduling operations. Based on the mechanism of the flywheel in cranes, we introduce the concept of an `energy buffer'. The stored energy will be depleted after the crane moves a certain distance. Let this distance be the energy buffer, and we assume that the energy buffer remains constant during the entire scheduling process. Energy will be temporarily stored in the energy buffer when the crane is lowering a container. Conversely, the stored energy will be reused when the crane is lifting a container, provided the energy has not dissipated. 

\vspace{0.3cm}

\noindent
\textbf{Paper Organization.} In this paper, we consider the crane scheduling problem with energy saving, which is crucial during the transporting phase of maritime logistics. The organization of the paper is as follows: In Section~\ref{sec2:formulation}, we describe our problem using a motivational example and present a mathematical formulation. In Section~\ref{sec3:eulerization}, we begin with an Eulerian perspective to analyze the problem under the condition of zero energy buffer $e = 0$, and generalize the analysis to the case where $e > 0$, and then propose an additive approximation algorithm based on the analysis.  In Section~\ref{sec4:dp}, we propose a DP approach that solves the crane scheduling problem in polynomial time, assuming the energy buffer and processing lengths are bounded by constants. In Section~\ref{sec5:arbitrary}, we adopt a Hamiltonian perspective to address the general case of the problem. Section~\ref{sec6:conclusion} concludes the paper with a summary and suggestions for future work.

\section{Problem Formulation}
\label{sec2:formulation}

We focus on a single container storage area with several containers to be scheduled. The layout of the storage area is two-dimensional, as shown in Figure \ref{layout}. Movements of the gantry crane can be performed in two directions: either to the seaside handover area or to the landside handover area. Due to the layout of the storage area, movements along the short side of the storage area are negligible compared to those along the long side. A gantry crane equipped with specific energy storage devices, as proposed in \cite{flynn2008saving}, will process a batch of containers. Note that the energy devices only store energy temporarily. The stored energy will be released after the crane moves a certain distance. Subsequently, if the crane needs to lift other containers, it will consume additional energy. We refer to this distance as the `energy buffer'. The device stores energy when a container is being lowered, and this energy can be reused for lifting operations within the energy buffer. Containers share the same size and weight, but they have different origins and destinations. Therefore, the movement of each container on the layout is modeled as a job.

Let the one-dimensional layout of the storage area be represented by $[1, \eta]$, where $\eta$ represents the total number of slots. Jobs will require moving containers between these discrete slots. The input consists of a set $J$ of $n$ jobs and an energy buffer $e$. Each job $j \in J$ has an origin slot $s_j$ and a destination slot $t_j$, where $s_j,t_j\in\{1,2,\cdots,\eta\}$. We define $l_j=|s_j-t_j|$ as the processing length of job $j$.

The objective of our scheduling problem is to determine a processing order for all jobs that minimizes total energy consumption. Energy consumption occurs during the lifting of containers. For any two consecutive jobs $j_i$ and $j_{i+1}$ in the processing order, if the distance between the destination of $j_i$ and the origin of $j_{i+1}$ does not exceed the energy buffer (i.e., $|s_{j_{i+1}}-t_{j_{i}}|\le e$), the crane can use stored energy from the previous lowering operation, thereby saving energy. Otherwise, the crane must consume extra energy to lift the container for $j_{i+1}$, as energy is released before the crane reaches the origin of $j_{i+1}$.

\begin{figure}[ht]
\centering
\captionsetup{justification=centering}
	\includegraphics[width=0.8\textwidth]{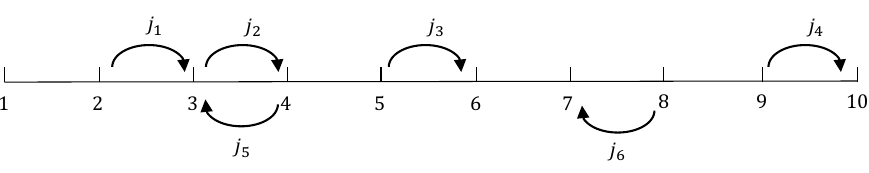}
	\caption{Example of job information} 
	\label{job_info_fig}
\end{figure}

Consider the example shown in Figure \ref{job_info_fig} with energy buffer $e=1$. If job $j_4$ is scheduled right after job $j_3$, the distance between $s_4=9$ and $t_3=6$ is greater than the energy buffer. Therefore, the crane cannot use the stored energy from $j_3$ and has to consume extra energy to lift $j_4$ at slot $9$. Traditionally, the gantry crane prefers unidirectional scheduling, instead of going back and forth \cite{bierwirth2009fast}. Following the traditional route: $j_1, j_2, j_3, j_4, j_6, j_5$, the total energy consumption is $4$, whereas the optimal solution consumes $3$ units of energy (with the processing route: $j_1, j_5, j_2, j_3, j_6, j_4$).

\section{Relationship to semi-Eulerization}
\label{sec3:eulerization}

In this section, we show the relationship between our problem and the semi-Eulerization. We begin with the special case where the energy buffer $e=0$, and demonstrate that the problem under this condition can be solved in linear time using the algorithm for finding a solution that adds the minimum number of edges in the semi-Eulerization problem. We then generalize the reduction to the case $e > 0$ and establish the connection between our problem and semi-Eulerization. Finally, we propose an additive approximation algorithm that combines a greedy method with solution enumeration.

\subsection{Crane Scheduling with Zero Energy Buffer}

Assume that the order of processing jobs in some solution is $j_1,j_2,\cdots,j_n$. When the energy buffer $e=0$, a job $j_{i+1}$ will not consume additional energy if and only if $t_{j_{i}} =s_{j_{i+1}}$, i.e., the destination of the previous job is precisely the origin of this job. Thus, our target can be naturally rephrased as maximizing the number of pairs of adjacent jobs $(j_{i},j_{i+1})$ such that the destination of $j_{i}$ is exactly the origin of $j_{i+1}$. 

Let $P=\{s_j, t_j|j\in J\}$. We build a graph $G$ consisting of $|P|$ vertices and $n$ edges, where each vertex corresponds to a slot in $P$ (we shall write vertex $x$ as the vertex representing slot $x$). For each job $j$, the graph has an edge from the vertex $s_j$ to the vertex $t_j$. 

Recall that a directed graph $G$ is Eulerian if an Euler circuit exists in $G$. The Eulerization is the process of adding edges to a graph to make it Eulerian. The semi-Eulerization is the process of adding edges to a graph to create an Euler path in $G$. Let $f(G)$ be the minimum number of edges to be added to semi-Eulerize $G$. 

\begin{lemma}\label{lemma1}
The minimum energy consumption under $e=0$ equals $f(G) + 1$. 
\end{lemma}

\begin{lemma}\label{lemma2}
Let $in_G(x)$ and $out_G(x)$ represent the indegree and outdegree of vertex $x$ in the graph $G$, respectively. We have
$$
f(G)=\frac{1}{2}\sum_{x\in P} \lvert in_G(x)-out_G(x)\rvert +(\# \text{ of Eulerian weakly connected components in }G) - 1
$$
\end{lemma}

See the constructive proofs in Appendix \ref{appB:proof1} and Appendix \ref{appC:proof2}. Thus, the optimal solution that adds the minimum number of edges to semi-Eulerize the given graph can be found in linear time. Based on the constructive proof of Lemma \ref{lemma1}, we can also transform the solution between the Optimal semi-Eulerization problem and our problem in linear time.

\subsection{Extend to the General Cases with Arbitrary Energy Buffer}\label{sec3.2}

In general cases where $e > 0$, we adopt the previous idea and construct a two-level graph to illustrate the connection between our problem and the semi-Eulerization problem. 

Let $P=\{s_j, t_j|j\in J\}$. We build a graph $G$ consisting of $2|P|$ vertices and $n$ edges: for each slot $x\in P$, there are two vertices $a_x$ and $b_x$ in $G$; for each job $j$, there is an edge from $a_{s_j}$ to $b_{t_j}$ in $G$. We refer to the vertices $\{a_x\}$ as the upper-level vertices and the vertices $\{b_x\}$ as the lower-level vertices.

\begin{definition}
Auxiliary Edges. A set of auxiliary edges $\mathcal A=\{(u_1,v_1), (u_2,v_2),\cdots,(u_n,v_n)\}$ is feasible if and only if it satisfies all the followings: (1) $|\mathcal A|=n$; (2) After adding all the auxiliary edges, each vertex $b_x$ from the lower level has equal indegree and outdegree; (3) For each edge $(u_i,v_i)$, $u_i$ is from the lower level and $v_i$ is from the upper level. Assume that $u_i=b_x$ and $v_i=a_y$, then $|x-y|\le e$.
\end{definition}

We use a feasible set $\mathcal A$ of auxiliary edges to balance the degree of the lower-level vertices, while these edges do not consume additional energy (i.e., the distance between the two ends of each auxiliary edge does not exceed the energy buffer). After adding all the auxiliary edges in $\mathcal A$, we remove all the isolated vertices, i.e., vertices without any adjacent edges. Let $G(\mathcal A)$ be the graph formed after processing all the above operations. 

\begin{definition}
Penalty Edges. Penalty edges are the edges of a minimum cardinality set of edges between upper-level vertices that semi-Eulerize $G(\mathcal A)$, i.e., $f(G(\mathcal A))$ equals the number of penalty edges, and the graph after adding these penalty edges is semi-Eulerian.
\end{definition}

\begin{figure}[h]
\centering
\captionsetup{justification=centering}
\includegraphics[width=\textwidth]{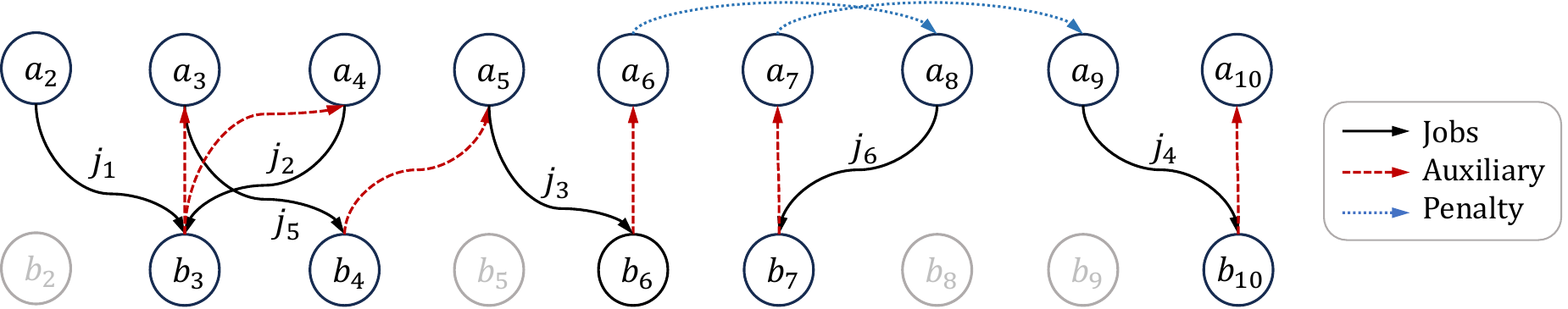}
\caption{Illustration of the two-level graph with auxiliary edges and penalty edges corresponding to the optimal solution for the example in Figure \ref{job_info_fig} with energy buffer $e = 1$} 
\label{job_info_twolevel_fig}
\end{figure}

We illustrate the two-level graph with auxiliary edges and penalty edges in Figure~\ref{job_info_twolevel_fig}, corresponding to the optimal solution of the example in Figure~\ref{job_info_fig} with energy buffer $e = 1$, whose optimal job sequence is $(j_1, j_5, j_2, j_3, j_6, j_4)$. Job $j_1$ has origin $s_{j_1}=2$ and destination $t_{j_1}=3$, corresponding to the original edge $(a_2,b_3)$ in black, and the same goes for other jobs. For adjacent pairs $(j_1,j_5),(j_5,j_2)$, and $(j_2,j_3)$ in the optimal sequence, since the distance between the destination of the first job and the origin of the second job does not exceed the energy buffer, only a red auxiliary edge is added. For adjacent pairs $(j_3,j_6)$ and $(j_6,j_4)$, since the distance between the destination of the first job and the origin of the second job exceeds the energy buffer, we first add a red auxiliary edge, then add a blue penalty edge. In conclusion, the total energy consumption for this example is $3$, which equals the number of penalty edges plus the one unit of energy consumption required to start the process.

\begin{figure}[h]
\centering
\captionsetup{justification=centering}
\includegraphics[width=0.92\textwidth]{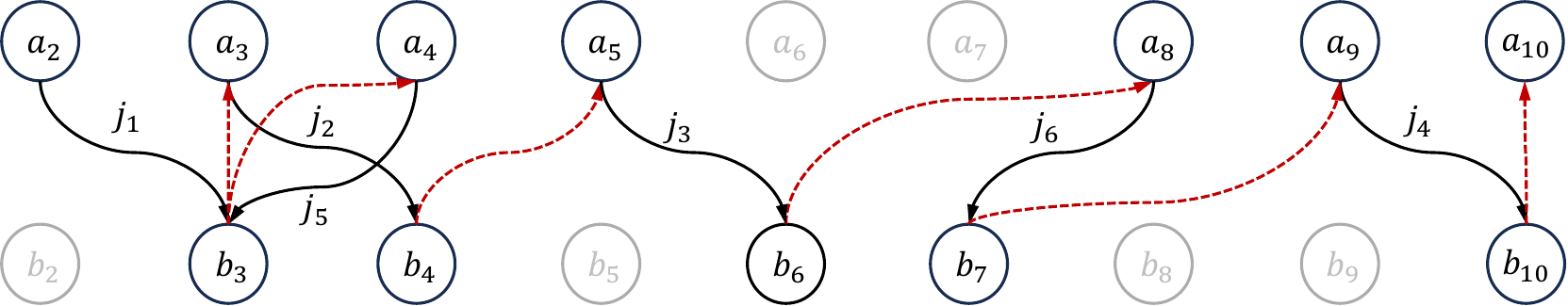}
\caption{Illustration of the two-level graph with auxiliary edges and penalty edges corresponding to the optimal solution for the example in Figure \ref{job_info_fig} with energy buffer $e = 2$} 
\label{job_info_twolevel_e2_fig}
\end{figure}

Considering the jobs in the example in Figure~\ref{job_info_fig}, when the energy buffer $e=2$, the optimal job sequence is $(j_1, j_5, j_2, j_3, j_6, j_4)$. Figure \ref{job_info_twolevel_e2_fig} depicts the corresponding two‐level graph. The optimal solution will only consume $1$ unit of energy to start the process. Thus, we can save energy between all the adjacent jobs in the optimal sequence; i.e., we can represent all the movements of the crane as auxiliary edges.

\begin{lemma}\label{lemma-general}
The minimum energy consumption is equal to $\displaystyle \min_{\mathcal A\text{ is feasible}} f(G(\mathcal A)) + 1$.
\end{lemma}

See the constructive proofs in Appendix \ref{appE:proof5}. Based on the constructive proof of Lemma \ref{lemma-general}, we can transform the solution between the optimal semi-Eulerization problem and our problem in linear time if we know the optimal set of auxiliary edges $\mathcal A^\star$. We will show in the next section how to determine $\mathcal A^\star$.

\subsection{Additive Approximation Algorithm}

We first propose a greedy algorithm to optimize the first part of $f(G(A))$ in $O(n\cdot \min(n,e))$-time, i.e., choose endpoints for all auxiliary edges to minimize $\frac{1}{2}\sum_{x\in P} |in_{G(A)}(x)-out_{G(A)}(x)|$. We first sort all the auxiliary edges in increasing order of their startpoints and then determine the endpoints following this order. For each auxiliary edge starting at vertex $b_x$, we try to make its endpoint a vertex $a_y$ such that $y\in [x-e,x+e]$ and its indegree is larger than its outdegree currently. If there are multiple choices, we choose the leftmost one. If such a vertex does not exist, we make its endpoint an arbitrary vertex $a_y$ such that $y\in [x-e,x+e]$. We enumerate at most $O(\min(n,e))$ vertices for each auxiliary edge. Therefore, the algorithm runs in $O(n\cdot\min(n,e))$ time. One can easily prove the optimality of the algorithm as any optimal solution can be transformed to a solution of this algorithm by swapping and moving endpoints of auxiliary edges.

\begin{algorithm}[htbp]
\captionsetup{justification=centering}
\caption{Additive Approximation Algorithm}\label{alg:dfs}
\begin{algorithmic}[1]
\Function{DFS}{$t,A$}
    \If{$t = k + 1$}
    	\State $temp \gets f(G(A\cup\{\textit{result of the greedy algorithm on remaining auxiliary edges}\}))$
        \State $answer \gets \min(answer, temp)$
        \State \Return
    \EndIf
    \ForAll{$i$ such that the endpoint of the $i$-th auxiliary edge has not been chosen}
        \For{$y \gets x - e$ \textbf{to} $x + e$} ~~~~// the startpoint of the $i$-th auxiliary edge is $b_x$
            \State $A \gets A\cup\{(b_x,a_y)\}$ ~~~// set the endpoint of the $i$-th auxiliary edge to be $a_y$
            \State \Call{DFS}{$t + 1, A$}
            \State $A \gets A\setminus\{(b_x,a_y)\}$
        \EndFor
    \EndFor
\EndFunction
\end{algorithmic}
\end{algorithm}

\begin{theorem}\label{theorem-general}
Algorithm \ref{alg:dfs} can find an approximate solution with additive error at most $n-k$ in $O(n^{k+1}\min(n,e)^{k+1})$ running time, for any positive integer $k\in[1,n]$.
\end{theorem}

In Algorithm \ref{alg:dfs}, we start by searching the endpoints of $k$ auxiliary edges in the graph, which yields a total of $O(n^k\min(n,e)^k)$ possibilities. Combined with the running time of the greedy algorithm, the running time of Algorithm \ref{alg:dfs} is $O(n^{k+1}\min(n,e)^{k+1})$.

 Consider the weakly connected components in the original graph; auxiliary edges would connect some of them in the optimal solution and the solution by our algorithm. Assume that there are $k'$ auxiliary edges that reduce the number of connected components in the optimal solution. In our algorithm, we have determined $k$ edges that reduce the number of connected components. 

\begin{itemize}
\item If $k'\leq k$, we have enumerated at least one possible case in which all of the $k'$ edges are included among the $k$ edges, and the remaining $k-k'$ edges are all from the optimal solution. For the remaining $n-k$ edges, applying the greedy algorithm guarantees that we minimize $\frac{1}{2}\sum_{x\in P} |in_{G(A)}(x)-out_{G(A)}(x)|$. Thus, we obtain the optimal solution.

\item Otherwise, $k'>k$, we have enumerated at least one possible case in which the $k$ edges are a subset of the $k'$ edges. Similarly, for the remaining $n-k$ edges, we apply the greedy algorithm to minimize $\frac{1}{2}\sum_{x\in P} |in_{G(A)}(x)-out_{G(A)}(x)|$. As we only need to determine the remaining $n-k$ endpoints, and each new endpoint can increase the number of Eulerian components by at most one, the additive error from our solution to the optimal solution is at most $n-k$. 
\end{itemize}

Therefore, Algorithm \ref{alg:dfs} returns a solution with an additive error of at most $n-k$.

\section{When Energy Buffer and Processing Lengths are Limited}
\label{sec4:dp}

In this section, we consider the case when energy buffer and processing lengths are bounded. We propose an exact dynamic programming approach to solve the crane scheduling problem in polynomial time when the bound is constant. Let $k=\max\{e, l_1,l_2,\cdots,l_n\}$ be the upper bound of energy buffer and processing lengths. Specifically, we have

\begin{theorem}\label{theorem1}
There exists a dynamic programming approach that can find the optimal solution to the crane scheduling problem in $O(n^{4k}k^{O(k)})$ running time.
\end{theorem}

The high-level idea is to find the optimal feasible set of auxiliary edges on the two-level graph from left to right, while maintaining the connectivity of the prefix of vertices with small cost in time and memory. We start from the unit-length case, where we can use a few entries to record the information we need. Then, we extend our approach to the general case by maintaining a disjoint-set data structure in the DP states. 

\subsection{Dynamic Programming Approach for the Unit-length Case}

We start from the unit-length case, when all the processing lengths of jobs and energy buffer are unit, i.e., $e=l_j=1, \forall j\in J$. In this case, we propose a DP approach that can solve the problem in $O(n^4)$ running time.

We first build the two-level graph according to the description in Section \ref{sec3.2} with only original edges (jobs). Then, we perform a left-to-right DP approach and add auxiliary edges and penalty edges along the prefix we considered in the states.

Let $a_{\le i}$ represent the set of all the vertices $a_x$ such that $x\in P$ and $x\le i$. Let $b_{\le i}$ represent the set of all the vertices $b_y$ such that $y\in P$ and $y\le i$. Also, we define $a_{\ge i}$ and $b_{\ge i}$ in the same way. For now, we assume that all the time slots in $P$ are adjacent, and we will show how to relax this assumption later.

\begin{definition}\label{def-unitdp}
Let $f(i,c_i,\gamma_{i,0},\gamma_{i,1},\delta_{i,0}, \delta_{i,1})$ represent the minimum energy consumption when only considering jobs related to vertices $a_{\le i+1}$ and $b_{\le i}$, where $c_i,\gamma_{i,0},\gamma_{i,1}\in\{\text{True},\text{False}\},$ $\delta_{i,0}\in\{0,1,\cdots,in_G(b_{i-1})+in_G(b_{i})+in_G(b_{i+1})\}, \delta_{i,1}\in \{0,1,\cdots,in_G(b_{i})+in_G(b_{i+1})+in_G(b_{i+2})\}$ on the condition that after adding all the auxiliary edges from $b_{\le i}$,

\begin{enumerate}[label= \roman*.)]
\item $c_i=\text{True}$ if vertices $a_i$ and $a_{i+1}$ are weakly connected, otherwise $c_i=\text{False}$.
\item $\gamma_{i,0}=\text{True}$ if all the vertices in $a_{\le i-1}$ that are weakly connected to $a_i$ are degree balanced, otherwise $\gamma_{i,0}=\text{False}$.
\item $\gamma_{i,1}=\text{True}$ if all the vertices in $a_{\le i-1}$ that are weakly connected to $a_{i+1}$ are degree balanced, otherwise $\gamma_{i,1}=\text{False}$ . 
\item the indegree of the vertex $a_i$ is $\delta_{i,0}$.
\item the indegree of the vertex $a_{i+1}$ is $\delta_{i,1}$.
\end{enumerate}
\end{definition}

Under the unit-length case, if we have fixed all the edges (including auxiliary edges and penalty edges) related to $a_{\le i+1}$ and $b_{\le i}$, the vertex set of a connected component will not change in any later steps if it does not include $a_{i}$ and $a_{i+1}$, since all new auxiliary edges can only attach to $a_{\ge i}$. This inspires us to only maintain information about at most two weakly connected components, such that they include the vertices $a_{i}$ and $a_{i+1}$, respectively. The vertex set of the remaining weakly connected components in the final two-level graph is either determined or has not yet appeared in the prefix we considered.

For transitions, we enumerate all the possibilities of the arrangement of all the auxiliary edges from vertex $b_{i+1}$. As the energy buffer is $1$ in the unit-length case, there are only $3$ different types of auxiliary edges from $b_{i+1}$: towards $a_{i},a_{i+1}$ and $a_{i+2}$. Thus each possible arrangement can be described as three integers $\Delta_{-1},\Delta_0,\Delta_1$ such that $\Delta_{-1}+\Delta_0+\Delta_1=in_G(b_{i+1})$ and there will be $\Delta_{-1}$ auxiliary edges from $b_{i+1}$ to $a_i$, $\Delta_{0}$ auxiliary edges from $b_{i+1}$ to $a_{i+1}$, and $\Delta_{1}$ auxiliary edges from $b_{i+1}$ to $a_{i+2}$. 

After adding these auxiliary edges, let $E_{i+1}$ be the set of all the original edges and auxiliary edges connected with $a_{i},a_{i+1},$ $a_{i+2},b_i,b_{i+1},b_{i+2}$. Then we determine the state $f(i+1,c_{i+1},\gamma_{i+1,0},\gamma_{i+1,1},\delta_{i+1,0}, \delta_{i+1,1})$  as follows:

\begin{itemize}
\item $c_{i+1}=\text{True}$ if and only if (i) $a_{i+1}$ and $a_{i+2}$ are weakly connected by edges in $E_{i+1}$, or (ii) $a_{i}$ and $a_{i+2}$ are weakly connected by edges in $E_{i+1}$ and $c_i=\text{True}$ holds simultaneously. Otherwise $c_{i+1}=\text{False}$.

\item $\gamma_{i+1,0}=\text{True}$ if and only if $\gamma_{i,1}=\text{True}$ and one of the following holds: (i) $c_i=\text{False}$ and $a_{i+1}$ is not weakly connected with $a_i$ by edges in $E_{i+1}$; (ii) $c_i=\text{False}$ and $a_{i+1}$ is weakly connected with $a_i$ by edges in $E_{i+1}$, and $a_{i}$ becomes balanced on degrees (i.e., $\delta_{i,0}+\Delta_{-1} = out_G(a_i)$); (iii) $c_i=\text{True}$ and  $a_{i}$ becomes balanced on degrees. Otherwise $\gamma_{i+1,0}=\text{False}$.

\item $\gamma_{i+1,1}=\text{True}$ if and only if one of the following holds: (i) $c_{i+1}=\text{False}$ and $a_{i+2}$ is not weakly connected with $a_i$ by edges in $E_{i+1}$; (ii) $c_{i+1}=\text{False}$ and $a_{i+2}$ is weakly connected with $a_i$ by edges in $E_{i+1}$, and $a_{i}$ becomes balanced on degrees; (iii) $c_{i+1}=\text{True}$ and $\gamma_{i+1,0} = \text{True}$. Otherwise $\gamma_{i+1,1}=\text{False}$.

\item $\delta_{i+1,0}=\delta_{i,1}+\Delta_{0}$ represents the indegree of $a_{i+1}$.

\item $\delta_{i+1,1}=\Delta_{1}$ represents the indegree of $a_{i+2}$, as no other auxiliary edges are added toward $a_{i+2}$ beforehand. 

\end{itemize}

\begin{figure}[h]
\centering
\captionsetup{justification=centering}
\includegraphics[width=0.95\textwidth]{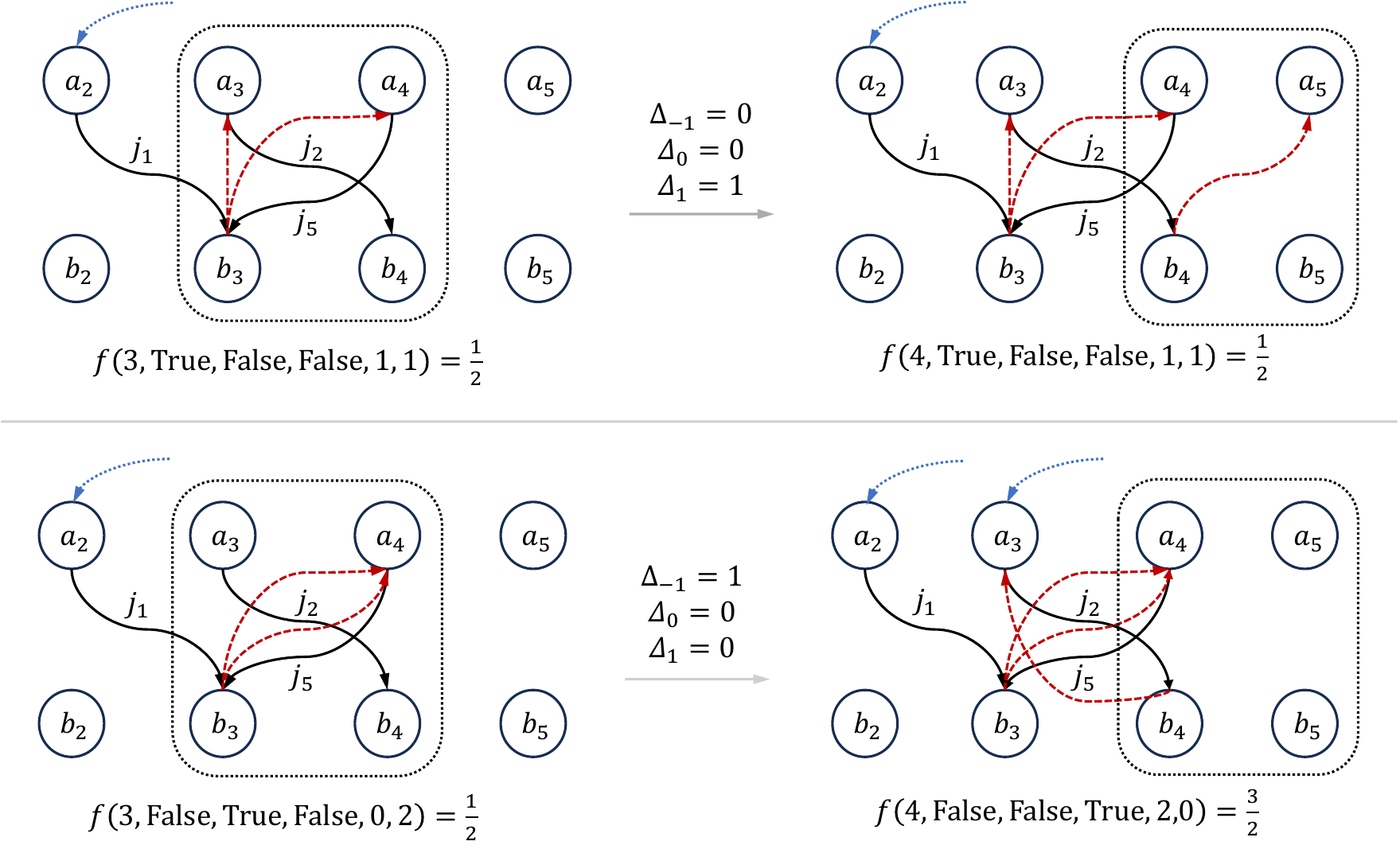}
\caption{Illustration of two possible transitions for the example in Figure \ref{job_info_fig}.} 
\end{figure}

Now we consider the contribution to $f(G(A))$ we can currently determine. After determining $\Delta_{-1},\Delta_0,\Delta_1$, the set of edges connected with $a_{i}$ in $G(A)$ will also be determined. Thus we can determine the contribution from $a_i$ to $f(G(A))$ by now as follows:

\begin{itemize}
\item the indegree and outdegree of $a_i$ are $\delta_{i,0}+\Delta_{-1}$ and $out_G(a_i)$, thus the contribution to the first part of $f(G(A))$ is $\frac{1}{2}|out_G(a_i)-\delta_{i,0}-\Delta_{-1}|$.

\item if $c_i=\text{False}$ and $a_i$ is not weakly connected with $a_{i+1}$ and $a_{i+2}$ by the original edges and auxiliary edges connected with $a_{i},a_{i+1},a_{i+2},b_i,b_{i+1},b_{i+2}$, then the composition of the weakly connected components which $a_i$ belongs to is also determined. This weakly connected component is Eulerian if and only if $\gamma_{i,0}=\text{True}$ and $a_{i}$ becomes balanced on degrees, and in this case, there will be one additional contribution to the second part of $f(G(A))$. 

\end{itemize}

Thus, we determine the transition equation as 

\begin{equation}
\begin{aligned}
f(i+1,&\,c_{i+1},\gamma_{i+1,0},\gamma_{i+1,1},\gamma_{i+1,0},\gamma_{i+1,1})\\
&= \min_{\substack{c,\gamma_{i,0},\gamma_{i,1},\delta_{i,0},\delta_{i,1},\\
\Delta_{-1},\Delta_0,\Delta_1}}
   \Bigl\{
     f(i,c_i,\gamma_{i,0},\gamma_{i,1},\delta_{i,0},\delta_{i,1})
     + \tfrac12\lvert\mathrm{out}_G(a_i)-\delta_{i,0}-\Delta_{-1}\rvert\\
&\qquad\quad
     + \mathbf{1}\{\text{the weakly connected component containing }a_i\text{ is Eulerian}\}
   \Bigr\}.
\end{aligned}
\label{chap4:dp:transition}
\end{equation}

We initialize $f(1, c_1=False, \gamma_{1,0}=False,\gamma_{1,1}=False,\delta_{1,0}=0,\delta_{1,1}=0)=1$. We enumerate the slot $i\in P$ from small to large, for every feasible state $(i,c_i,\gamma_{i,0},\gamma_{i,1},\delta_{i,0},\delta_{i,1})$ and every valid split $\Delta_{-1}+\Delta_{0}+\Delta_{1}=in_{G}(b_{i+1})$, we update $f(i+1,c_{i+1},\gamma_{i+1,0},\gamma_{i+1,1},\gamma_{i+1,0},\gamma_{i+1,1})$ according to trainsition equation (\ref{chap4:dp:transition}). We repeat this process until $i$ becomes the largest slot.

We now show how to obtain the optimal solution. We enumerate all the DP states on the last slots $f(n,c_n,\gamma_{n,0},\gamma_{n,1},\delta_{n,0},\delta_{n,1})$ and add the contribution from the last slot to $f(G(\mathcal A))$. As all the auxiliary edges are determined, the contribution of $a_i$ can be calculated in the same way as transitions. Thus we have

$$
\begin{aligned}
opt
&= \min_{c_n,\gamma_{n,0},\gamma_{n,1},\delta_{n,0},\delta_{n,1}}
   \Bigl\{
     f(n,c_n,\gamma_{n,0},\gamma_{n,1},\delta_{n,0},\delta_{n,1})
     + \tfrac12\lvert\mathrm{out}_G(a_n)-\delta_{n,0}\rvert\\
&\qquad\quad
     + \mathbf{1}\{\text{the weakly connected component containing }a_n\text{ is Eulerian}\}
   \Bigr\}.
\end{aligned}
$$

\noindent
{\textbf{Correctness.}} The dynamic programming approach returns the smallest energy consumption since it tries all the feasible sets of auxiliary edges. We can also obtain the corresponding feasible set $\mathcal{A}^\star$ of auxiliary edges by recording the optimal transition for each state, and performing backtracking from the states leads to $opt$ on the last slot. Then, based on the constructive proof of Lemma \ref{lemma-general}, we can obtain the optimal processing order of all the jobs.

\vspace{0.3cm}

\noindent
{\textbf{Time Complexity.}} The total running time can be represented as 
$$
\begin{aligned}
& O(1)\times \sum_{i\in P}\sum_{\delta_0}\sum_{\delta_1}(in_G(b_{i+1}))^2 \\
=& O(1)\times \sum_{i\in P}(in_G(b_{i+1}))^2 \bigg(1 + \sum_{j\in\{i-1,i,i+1\}} in_G(b_{j})\bigg)\bigg(1 + \sum_{j\in\{i,i+1,i+2\}} in_G(b_{j})\bigg)\\
\le & O(n^2)\times \sum_{i\in P}(in_G(b_{i+1}))^2 \le O(n^2)\times O(n^2)=O(n^4)
\end{aligned}
$$

The last inequality holds since $\sum_{x_i\ge 0, x_1+x_2+\cdots+x_n=n}x_i^2\le n^2$. Thus, our proposed approach can solve the unit-length case in $O(n^4)$ time.

\subsection{Extended Dynamic Programming Approach}

Now we extend our approach to a more general case when energy buffer and processing lengths are bounded by a parameter $k$. To avoid complex case discussions on connectivity and whether each weakly connected component is Eulerian, we propose the concept of configuration, which efficiently maintains all the necessary information for transition.

\begin{definition}\label{def-configuration}
We define the configuration $\mathcal C_i$ of a certain slot $i$ to be a combination of all the following:
\begin{enumerate}
\item A disjoint-set data structure  $\mathcal U_i$ maintains whether each pair of vertices among the vertices $a_{i-k+1},a_{i-k+2},\cdots,a_i,a_{i+1},\cdots,a_{i+k}$ is weakly connected. It supports updating the connectivity information in $O(\alpha(k))$ time when adding an edge between these vertices. 
\item A boolean vector $\mathcal G_i=\langle\gamma_{i,1-k},\gamma_{i,2-k},\cdots,\gamma_{i,-1},\gamma_{i,0},\gamma_{i,1},\cdots,\gamma_{i,k}\rangle$ where $\gamma_{i,x}=\text{True}$ identifies that all the vertices in $a_{\le i-k}$ that are weakly connected with $a_{i+x}$ are balanced, otherwise $\gamma_{i,x} = False$. Note that all the vertices that belong to the same weakly connected components share the same $\gamma$ value. Thus, it supports updating in $O(1)$ time when adding an edge connecting to these vertices.
\item An integer vector $\mathcal D_i=\langle\delta_{i,1-k},\delta_{i,2-k},\cdots,\delta_{i,-1},\delta_{i,0},\delta_{i,1},\cdots,\delta_{i,k}\rangle$ where $\delta_{i,x}\in\{0,1,$ $\cdots,\sum_{j=i+x-k}^{i+x+k}in_G(b_j)\}$ identifies the indegree of the vertex $a_{i+x}$. It supports updating in $O(1)$ time when adding an edge connecting to these vertices.
\end{enumerate}
\end{definition}

Notice that the number of distinct $\mathcal U_i$ is at most $(2k)^{2k}$, the number of distinct $\mathcal G_i$ is at most $2^{2k}$, and the number of distinct $\mathcal D_i$ is at most $n^{2k}$ as the total indegree caused by auxiliary edges is bounded by the the number of jobs $n$.  Thus, the number of distinct configurations $\mathcal C_i$ is bounded by $(2k)^{2k}\cdot 2^{2k}\cdot n^{2k}=O(n^{2k}k^{O(k)})$. 

We consider the changes in $\mathcal C_i$ when adding a set $E$ of auxiliary edges from the vertex $b_{i+1}$ and original edges connected with $a_{\le i+k+1}$ and $b_{i+1}$. We temporarily extend $\mathcal C_i$ by maintaining two additional vertices $a_{i+k+1}$ and $b_{i+1}$. Then we can add those edges one by one while maintaining all the information needed in $O(|E|)$ time. Moreover, as the energy buffer and processing lengths are bounded by $k$, the ends of all the edges in $E$ will be one of $a_{i-k+1},\cdots,a_{i+k+1}$ and $b_{i+1}$, thus there will be at most $O(k)$ different edges distinguished by their ends. We can do the update procedure for all the edges with the same ends simultaneously, then the running time can be reduced to $O(k)$.

After processing all the operations, we remove those variables representing the information about vertices $a_{i-k+1}$ and $b_{i+1}$. We use $\mathcal C_{i+1}=tr(\mathcal C_i,E)$ to represent the final configuration we get. Note that as the energy buffer and processing lengths are bounded by $k$, all information related to $a_{i-k}$ we mentioned above will never change once we determine all the auxiliary edges from vertices in $b_{\le i+1}$.

\begin{definition}\label{def-kdp}
Let $f(i,\mathcal C_i)$ represent the minimum energy consumption for only considering jobs related to vertices $a_{\le i+k}$ and $b_{\le i+1}$, on the condition that the graph $G$ with all the added auxiliary edges is compatible with the configuration $\mathcal C_i$.
\end{definition}

Similar to the unit-length case, for transitions, we consider enumerating all the possibilities of the arrangement of all the auxiliary edges from vertex $b_{i+1}$. There are only $2e+1$ different types of auxiliary edges from $b_{i+1}$: towards $a_{i+1-e},a_{i+1-e+1},\cdots,a_{i+1+e}$. Thus each possible arrangement can be described as a vector of $2e+1$ non-negative integers $\Delta=\langle\Delta_{-e},\cdots,\Delta_0,\cdots,\Delta_e\rangle$ such that $\sum_{-e\le x\le e}\Delta_{x}=in_G(b_{i+1})$ and there will be $\Delta_{x}$ auxiliary edges from $b_{i+1}$ to the vertex $a_{i+1+x}$ for all the $x\in[-e,e]$ if the vertex $a_{i+1+x}$ exists, otherwise $\Delta_{x}$ should be $0$. Then we create the set of edges $E(i+1,\Delta)$ consisting of all the auxiliary edges according to $\Delta$ and all the original edges ending at the vertex $b_{i+1}$. The state will transit to $f(i+1,tr(\mathcal C_i,E(i+1,\Delta))$ after adding these edges.

Now we consider the contribution to $f(G(A))$ we can currently determine as follows:
\begin{itemize}
\item the indegree and outdegree of $a_{i+1-k}$ are $\delta_{i,1-k} + \Delta_{-k}$ and $out_G(a_{i+1-k})$, thus the contribution to the first part of $f(G(A))$ is $\frac{1}{2}|out_G(a_{i+1-k})-\delta_{i,1-k} - \Delta_{-k}|$.
\item from the procedure of updating $\mathcal C_i$ by $E(i+1,\Delta)$ we can know whether $a_{i+1-k}$ is weakly connected with other vertices $a_{x}$ such that $x>i+1-k$. If not, then the weakly connected component which $a_{i+1-k}$ belongs to is Eulerian if and only if $\gamma_{i,1-k}=\text{True}$ and $a_{i+1-k}$ becomes balanced on degrees. In this case, there will be one additional contribution to the second part of $f(G(A))$. 
\end{itemize}

Thus, we determined the transition equation as 

\begin{equation}
\begin{aligned}
&f\bigl(i+1,\mathcal{C}_{i+1}=tr(\mathcal C_i,\,E(i+1,\Delta))\bigr) \\
&\quad\; = \min_{i,\mathcal C_i,\Delta} \Bigl\{
     f(i,\mathcal C_i)
     + \tfrac12\lvert out_G(a_{i+1-k})-\delta_{i,1-k}-\Delta_{-k}\rvert\\
&\quad\; + \mathbf{1}\{\text{the weakly connected component containing }a_{i+1-k}\text{ is Eulerian}\}
  \Bigr\}.
\end{aligned}
\label{chap4:dp:transition-ext}
\end{equation}

\begin{figure}[htbp]
\centering
\captionsetup{justification=centering}
\includegraphics[width=0.95\textwidth]{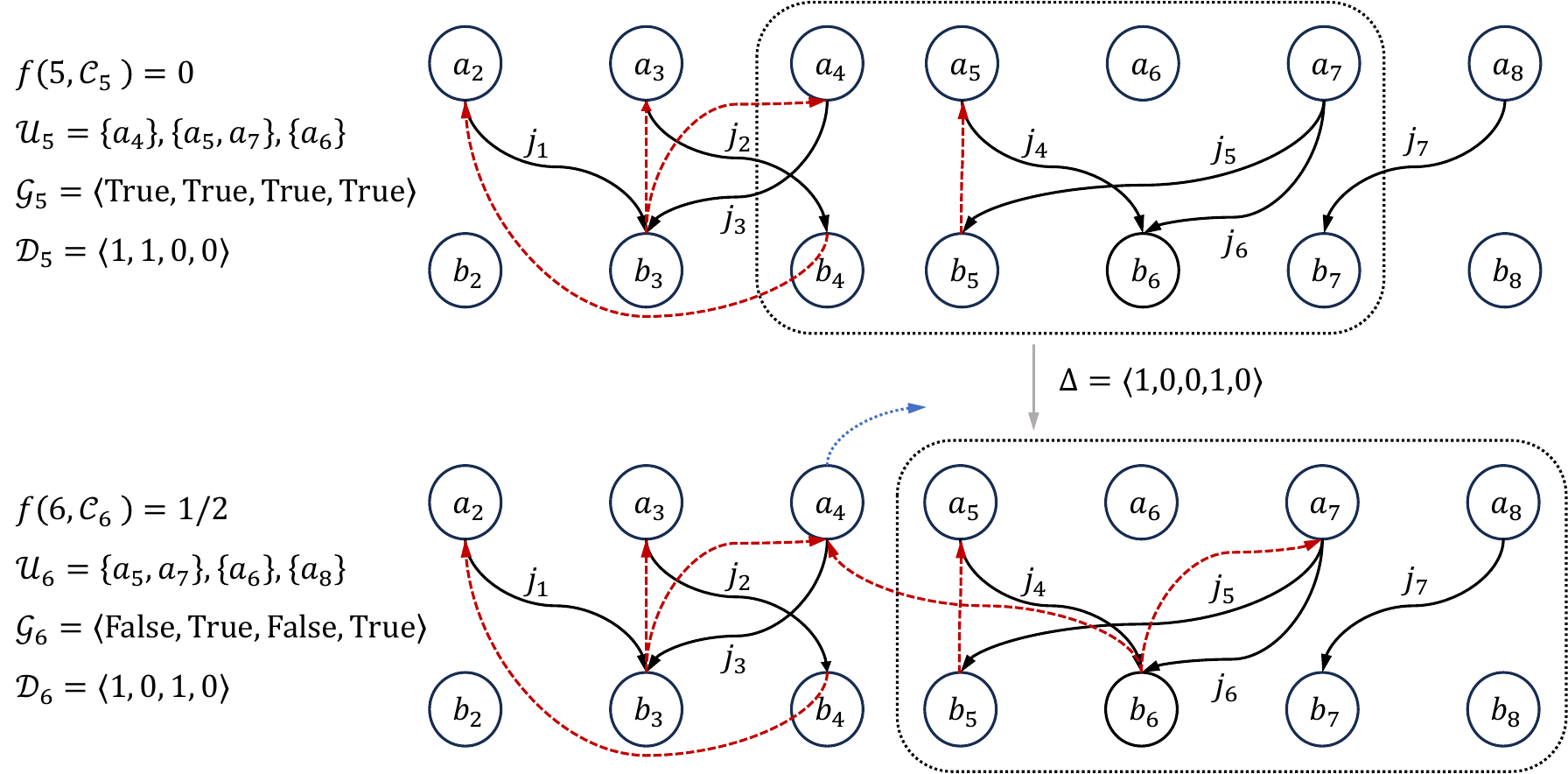}
\caption{Illustration of a possible transition for an instance with $k=2$.} 
\end{figure}

We eumerate the slot $i\in P$ from small to large, for every feasible state $(i,\mathcal{C}_i)$ and every valid split $\Delta$, we update $f(i+1,\mathcal{C}_{i+1})$ according to trainsition equation (\ref{chap4:dp:transition-ext}). We repeat this process until $i$ becomes the largest slot. 

Now we show how to get the optimal solution. We enumerate all the DP states on the last slots $f(n,\mathcal{C}_n)$ and add the contribution from the last $k$ slots to $f(G(\mathcal A))$. As all the auxiliary edges are determined, the contribution of $a_{n-k+1},a_{n-k+2},\cdots,a_n$ can be calculated in the same way as transitions. Thus we have

$$
\begin{aligned}
&opt = \min_{c,\mathcal{C}_n}
   \Bigl\{
     f(n,\mathcal{C}_n)
     + \sum_{n-k+1\le i\le n}\tfrac12\lvert\mathrm{out}_G(a_i)-\delta_{n,i-n}\rvert\\
& + (\#~\text{of Eulerian weakly connected components containing one of } a_{n-k+1},a_{n-k+2},\cdots,a_n)
   \Bigr\}.
\end{aligned}
$$

\noindent
{\textbf{Time Complexity.}} The total running time can be represented as

$$
O(k^{O(1)})\times \sum_{i\in P}\sum_{\mathcal{C}_i}(in_G(b_{i+1}))^{2e} \le  O(n^{2k}k^{O(k)})\times \sum_{i\in P}(in_G(b_{i+1}))^{2e} \le O(n^{2k+2e}k^{O(k)})
$$

Thus, our proposed approach can solve the problem in $O(n^{4k}k^{O(k)})$ time as $e\le k$. When $k$ is a fixed constant, our approach can solve the problem in polynomial time $O(n^{4k})$.

\vspace{0.3cm}

\noindent
{\textbf{Remove the Contiguity Assumption.}} 

\vspace{0.2cm}

If the slots are non-contiguous, let $\mathcal{P}=\{\rho_1,\cdots,\rho_m\}$ be the sorted positions. We relabel the two-level graph so that $a_i$ and $b_i$ correspond to slot $\rho_i$. We then apply the DP approach with the following modification. Define the neighbor set $\mathcal{N}_{i+1}=\{x\in[m]|\rho_x-\rho_{i+1}\leq e\}$. We enumerate $\{\Delta_x\}_{x\in \mathcal{N}_{i+1}}$ satisfying $\sum_{x\in\mathcal{N}_{i+1}}\Delta_x=in_G(b_{i+1})$ instead of enumerating vector $\Delta$. The remaining procedures remain unchanged.

\section{Crane Scheduling with Arbitrary Energy Buffer and Processing Lengths}
\label{sec5:arbitrary}

In this section, we discuss the crane scheduling problem with arbitrary energy buffers. We adopt a Hamiltonian perspective by reducing the original problem to a path cover problem. We then propose an exact algorithm based on DP, which runs in time $O(2^nn^2)$. Furthermore, for a tractable variant of the general problem, we show that it can be solved in polynomial time when it reduces to a path cover instance on acyclic interval digraphs.

\subsection{Transformation to Path Cover Problem}

\begin{definition}
Given a directed graph $D(V,E)$ (digraph for short), each vertex $v\in V$ has two interval sets $S_v$ and $T_v$, namely the source set and the terminal set. Given $u,v\in V$, $uv$ is an edge of $D$ if and only if $T_u\cap S_v\neq \emptyset$, where loops are allowed but multiple edges are not. Such a directed graph $D(V,E)$ is called an interval digraph \cite{drachenberg1994interval}.
\end{definition}

Each job can be regarded as a vertex in the interval digraph. The corresponding source set $S_j=\{s_j\}$, where $s_j$ is the origin of job $j$. The corresponding terminal set $T_j=[t_j-e,t_j+e]$, which is an interval. For any two vertices $u,v\in V$, there will be a directed edge $uv\in E$ if $T_u\cap S_v\neq \emptyset$. Therefore, the crane can schedule job $u$ and job $v$ in turn without consuming extra energy, since the origin $s_v$ is located in the terminal set of job $u$: $s_v\in [t_u-e,t_u+e]$.

\begin{definition}
Given a directed graph $D(V,E)$, the vertex-disjoint path cover problem is to minimize the number of vertex-disjoint elementary paths that cover all the vertices. Let the minimum number of vertex-disjoint paths be $k$, and let the set ${D_1,D_2,\ldots ,D_k}$ denote the vertex-disjoint path set. We have $D_i\cap D_j=\emptyset, \forall i,j\in {1,\cdots,k}, $ and $i\neq j$. Let $D_i=(V_i,E_i)$ for $\forall i \in{1,\cdots,k}$. Then ${D_1,D_2,\cdots,D_k}$ is a vertex-disjoint cover of digraph $D$ if ${V_1,\cdots,V_k}$ partitions $V$, i.e., $V=\bigcup_{i=1}^{k}{V_i}$ while $V_i\cap V_j=\emptyset, i\neq j$. Note that a zero-length path is allowed, but a circuit is not allowed \cite{boesch1977covering}.
\end{definition}

Various path cover problems are introduced in \cite{ntafos1979path,linial1978covering,ntafos1984computational}. Based on different settings and properties, the problem complexity is different. Finding the minimum number of vertex-disjoint path covers on acyclic digraphs is polynomially solvable since it can be reduced to the maximum matching problem. Minimization on the general directed graph is NP-complete \cite{boesch1977covering}, but the interval digraph case remains unknown. On the other hand, Arallah and Kosaraju considered the general line model of the stacker crane problem to minimize the total distance \cite{atallah1988efficient}. 

Consider the vertex-disjoint path cover solution on the interval digraph. For each path, we can process all the jobs along the path. Notice that only one unit of energy is required to start processing the first job on each path, and all the remaining jobs can use the energy saved from the previous job. Thus, the crane scheduling problem can be regarded as a vertex-disjoint path cover problem on the interval digraph.

\subsection{An Exact Dynamic Programming Algorithm}

Given the directed graph $D(V,E)$, let $\Gamma_i=\{j\in V: (j,i)\in E\}$ denote the set of predecessors of vertex $i$ in $D$. We define $f(\mathcal S,i)$ for $\mathcal S\subseteq V$ and $i\in \mathcal S$ be the minimum number of vertex-disjoint paths required to cover all the vertices in $\mathcal S$ and the last path is ending with vertex $i$. Our DP procedure is initialized by setting $f(\{i\},i)=1$ for each vertex $i\in V$. For any state $\mathcal S\subseteq V$ and any vertex $i\in \mathcal S$, we consider the structure of the path covering vertex $i$ and take the minimum over all possible $j\in S\setminus\{i\}$ which is the last vertex covered. A forward DP recursive function is applied.

$$
f(\mathcal S, i) = \min \left \{  
\begin{array}{ll}
f(\mathcal S\setminus\{i\},j), & j\in \Gamma_i \cap \mathcal{S},  \\
f(\mathcal S\setminus\{i\},j) + 1, &  j \in \mathcal{S} - \Gamma_i.
\end{array}
\right.
\label{arbi:DP}
$$

When $j\in \Gamma_i$, which means there is a directed edge $(j,i)\in E$ in the directed graph $D$. Then the current path ends at vertex $j$, and we can cover vertex $i$ immediately afterwards. On the other hand, when $j\notin \Gamma_i$, we need to start a separate path to cover the vertex $i$. To compute $f(\mathcal{S}, i)$, we take the minimum of all $j$. Note that the state becomes infeasible for $j\notin \mathcal{S}$, and we let $f(S,j)=+\infty$. The minimum number of paths that cover all the vertices is $\min_i f(J,i)$. Since there are $O(2^n)$ subsets and up to $n$ choices of last vertex for each, the DP has a table size of $O(2^n n)$. The transition of each state takes $O(n)$ time; thus, the DP runs in $O(2^nn^2)$ time.

\subsection{A Tractable Special Case: Path Cover on Acyclic Interval Digraphs}

If the interval digraph representing the given instance of crane scheduling is acyclic, we can solve the crane scheduling problem in polynomial time as follows. For an acyclic digraph $D(V,E)$, we create a graph $D^{'}(V^{'},E^{'})$. Each vertex $v\in V$ is replaced with a pair of vertices $x_v,y_v\in V^{'}$ such that given $uv\in E$, $x_uy_v\in E^{'}$. Therefore, all edges in $E^{'}$ are from one part to the other part, i.e., $V^{'}$ can be partitioned into two disjoint sets ${x_1,\cdots,x_n}$, ${y_1,\cdots,y_n}$ and no arc connects any two vertices in the same set. Thus, graph $D^{'}$ is bipartite.

\begin{lemma}
Let $\rho$ be the minimum number of vertex-disjoint paths that cover all vertices of the acyclic digraph $D(V,E)$. Let $\mu$ be the number of edges in the maximum matching of the corresponding bipartite graph $D^{'}(V^{'},E^{'})$. We have $\rho =|V|-\mu$.
\label{lemma5-2}
\end{lemma}

The proof of Lemma \ref{lemma5-2} is shown in Appendix \ref{appG:proof5-2}. Therefore, we can use the maximum matching algorithm (like the Hopcroft–Karp algorithm) to solve our problem. An illustrative example of the path-cover reduction on an acyclic digraph is given in Appendix \ref{appH:fig5-2}. 

\section{Conclusion}
\label{sec6:conclusion}

In this paper, we propose a novel mechanism for a single crane to schedule containers, aiming to reuse the energy from already lifted containers and minimize the total energy consumption of the entire scheduling plan. We establish a basic model considering a one-dimensional storage area and provide a systematic complexity analysis of the problem. We bridge the gap between the classic crane scheduling problems studied in the Operations Research area and their underlying theoretical foundations. We also introduce a paradigm that integrates both the Eulerian and Hamiltonian perspectives, providing a robust framework for addressing the problem.

Future research should extend the current approach to handle arbitrary-length jobs, transforming the problem into a path cover problem on general digraphs, and investigate the hardness of the general version of the problem. Additionally, further investigation is needed to consider more sophisticated energy consumption functions, including exploring various factors that impact energy usage, such as different types of cranes, varying container weights, and dynamic environmental parameters, including energy loss. Moreover, applying the proposed solutions to real-world crane scheduling scenarios will validate their practicality and effectiveness.

\bibliography{refs}

\newpage

\appendix

\noindent
\section{Proof of Lemma \ref{lemma1}}
\label{appB:proof1}

\begin{proof}
Let $opt$ be the minimum energy consumption for the original problem.

1. $opt\ge f(G) + 1$: Assume that the order for processing jobs corresponding to $opt$ is $j_1,j_2,\cdots,j_n$. For each pair of adjacent jobs $(j_{i}, j_{i+1})$, if $t_{j_{i}}\ne s_{j_{i+1}}$, then we add an edge in $G$ from the vertex $t_{j_{i}}$ to the vertex $s_{j_{i+1}}$. The number of edges added equals $opt-1$, and these edges semi-Eulerize the graph since an Euler path exists following the processing order of jobs. Thus $opt-1\ge f(G) \Rightarrow opt\ge f(G)+1$.

2. $opt\le f(G) + 1$: After adding the edges corresponding to $f(G)$, we can find an Euler path in the graph. If we ignore all the added edges, the Euler path will be split into several disjoint subpaths, and the number of subpaths is equal to the number of added edges $+1$, i.e., $f(G)+1$. For each subpath, we can process all the jobs corresponding to the edges in the subpath, in the same order as traversing the subpath. It will only consume one unit of energy to process all the jobs in each subpath, and all the jobs will be processed since we do not ignore any edge corresponding to a job. Thus $opt\le f(G)+1$.
\end{proof}
 
\section{Proof of Lemma \ref{lemma2}}
\label{appC:proof2}

\begin{proof}

Let $C_1, C_2, \cdots, C_k$ be the weakly connected components of the original graph $G$, and let $C_1, \cdots, C_{k^{'}}$ be those components that contain at least one degree imbalance of the vertex (i.e., $\exists x\in V(C_i),in_G(x) \neq out_G(x), i\in\{1,2,\cdots,k\}$). For the remaining components $C_{k'+1},\cdots, C_k$, all vertices are balanced (i.e., these components are Eulerian). For each imbalanced component $C_i, i\in\{1,2,\cdots,k'\}$, we choose two vertices $\alpha_i,\beta_i\in V(C_i)$ arbitrarily such that $in_{\alpha_i}<out_{\alpha_i},in_{\beta_i}>out_{\beta_i}$. For each remaining balanced components $C_{i}, i\in\{k'+1,k'+2,\cdots,k\}$, we choose an arbitrary vertex $\alpha_i=\beta_i\in V(C_i)$. 

We first connect all the components into a weakly connected graph $\hat{G}$ by adding $k-1$ directed edges $(\beta_1,\alpha_{2}),(\beta_2,\alpha_{3}),\cdots,(\beta_{k-1},\alpha_k)$. Then, we repeat the following procedure, until all the vertices are balanced except at most one vertex $\alpha$ such that its indegree is larger than its outdegree by $1$, and at most one vertex $\beta$ such that its indegree is smaller than its outdegree by $1$: Choose two vertices $\alpha,\beta$ such that the indegree of $\alpha$ is less than its outdegree in the current graph, and the outdegree of $\beta$ is less than its indegree in the current graph. Then add a directed edge from $\beta$ to $\alpha$ and update the degree information of the vertices.

Let $m$ be the total number of edges added in our construction. We have 

$$
\begin{array}{lll}
m &=&k-1+\frac{1}{2}\sum_{x\in V(\hat{G})} \lvert in_{\hat G}(x)-out_{\hat G}(x)\rvert-1\\ [10pt]
&=&\bigg(k'-1+\frac{1}{2}\sum_{x\in V(\hat{G})} \lvert in_{\hat G}(x)-out_{\hat G}(x)\rvert \bigg)+(k-k') -1\\ [10pt]
&=&\frac{1}{2}\sum_{x\in P} \lvert in_G(x)-out_G(x)\rvert \;+ \\ [10pt]
& &(\# \text{ of Eulerian weakly connected components in }G) - 1
\end{array}
$$

1. $m \geq f(G)$: By construction, after adding $m$ edges, the graph becomes weakly connected and all the vertices are balanced except at most one vertex $\alpha$ whose indegree is larger than its outdegree by $1$, and at most one vertex $\beta$ whose indegree is smaller than its outdegree by $1$. Thus, it satisfies the condition for finding an Eulerian path, i.e., the graph is semi-Eulerian.

2. $m \leq f(G)$: If all the weakly connected compenents $C_1,C_2,\cdots, C_k$ in $G$ are Eulerian, then it requires at least $k-1$ edges to connect them, which equals $m$. Otherwise, as each added edge can reduce imbalanced degree by at most $2$ (increase one indegree and one outdegree), to meet the condition of semi-Eulerian requires at least $\frac{1}{2}\sum_{x\in P}\lvert in_G(x) - out_G(x)\rvert - 1$ edges. Each of the remaining Eulerian weakly connected components requires one additional edge to connect it with the others. Therefore, any process that semi-Eulerizes $G$ requires adding at least $m$ edges.
\end{proof} 
\section{Proof of Lemma \ref{lemma-general}}
\label{appE:proof5}

\begin{proof}
Let $opt$ be the minimum energy consumption for the original problem.

1. $opt\ge\min_{\mathcal A\text{ is feasible}}  f(G(\mathcal A)) + 1$: Assume that the order for processing jobs corresponding to $opt$ is $j_1,j_2,\cdots,j_n$. For each pair of adjacent jobs $(j_{i-1}, j_i)$, if the distance between the destination of $j_{i-1}$ and origin of $j_i$ does not exceed the energy buffer (no additional energy consumption for processing job $j_i$), we only add an auxiliary edge from the vertex $b_{t_{j_{i-1}}}$ to the vertex $a_{s_{j_i}}$; otherwise, we add an auxiliary edge from the vertex $b_{t_{j_{i-1}}}$ to the vertex $a_{t_{j_{i-1}}}$, and a penalty edge from the vertex $a_{t_{j_{i-1}}}$ to the vertex $a_{s_{j_i}}$. In addition, we add one more auxiliary edge from the vertex $b_{t_{j_{n}}}$ to the vertex $a_{t_{j_{n}}}$. Let the set of auxiliary edges we add in this construction be $\mathcal A^\star$; it is straightforward that $\mathcal A^\star$ is feasible. The number of penalty edges (do not include the auxiliary edges) we add equals $opt-1$, and these edges semi-Eulerize the graph since there exists an Euler path following the processing order of jobs, starting form the vertex $a_{s_{j_1}}$, ending at the vertex $a_{t_{j_n}}$. Thus we have $opt-1\ge f(G(\mathcal A^\star)) \Rightarrow opt\ge \min_{\mathcal A\text{ is feasible}}f(G(\mathcal A))+1$.

2. $opt\le \min_{\mathcal A\text{ is feasible}}  f(G(\mathcal A)) + 1$: Assume that $\mathcal A^\star = \arg \min_{\mathcal A\text{ is feasible}}  f(G(\mathcal A))$. After adding the edges in $\mathcal A^\star$ and edges corresponding to $f(G(\mathcal A^\star))$, we can find an Euler path in the graph. Then we ignore all the penalty edges added corresponding to $f(G(\mathcal A^\star))$. The Euler path will be split into several disjoint subpaths, and the number of subpaths is equal to $f(G(\mathcal A^\star))+1$. Each subpath consists of original edges and auxiliary edges alternatively, and we can process all the jobs corresponding to the edges in the subpath in the same order as traversing the subpath. It will only consume one unit of energy to process all the jobs in each path, as all the auxiliary edges satisfy the energy buffer limit. And all the jobs will be processed since we do not ignore any edge corresponding to a job. Thus $opt\le f(G(\mathcal A^\star))+1=\min_{\mathcal A\text{ is feasible}} f(G(\mathcal A)) + 1$.
\end{proof} 
\section{Proof of Lemma \ref{lemma5-2}}
\label{appG:proof5-2}

\begin{proof}
Let $\omega$ be the number of edges in the minimum path coverage. We have $\rho =|V|-\omega$. Therefore, we only need to prove $\mu =\omega$.
Considering the initial case when $\mu =0$, no directed edges exist in $D$. Obviously $\rho =|V|-\mu =|V|-0=0$.
Based on the initial case, when we add one matching edge $x_i y_j$ to $D^{'}$, edge $ij$ will exist in the corresponding path cover. Therefore, $\rho$ is reduced by $1$. In this way, each time the number of matching edges increases by one, the path coverage decreases by one. We keep doing this until the number of matching edges could no longer increase, and then the number of path covers could no longer decrease.
On the other hand, for every edge $ij$ in the path cover, we have the corresponding matching edge $x_iy_j$ in the bipartite graph $D^{'}$. If there exist two edges $x_iy_j$ and $x_iy_k$ with $j\neq k$, $ij$ and $ik$ are in the path cover graph $D$, which is a contradiction.
As a result, $\mu$ and $\omega$ is a one-to-one correspondence.
\end{proof}
\section{Example of Arbitrary Length Case to Interval Acyclic Digraph}
\label{appH:fig5-2}

\begin{figure}[ht]
\centering
\captionsetup{justification=centering}
\includegraphics[width=0.8\textwidth]{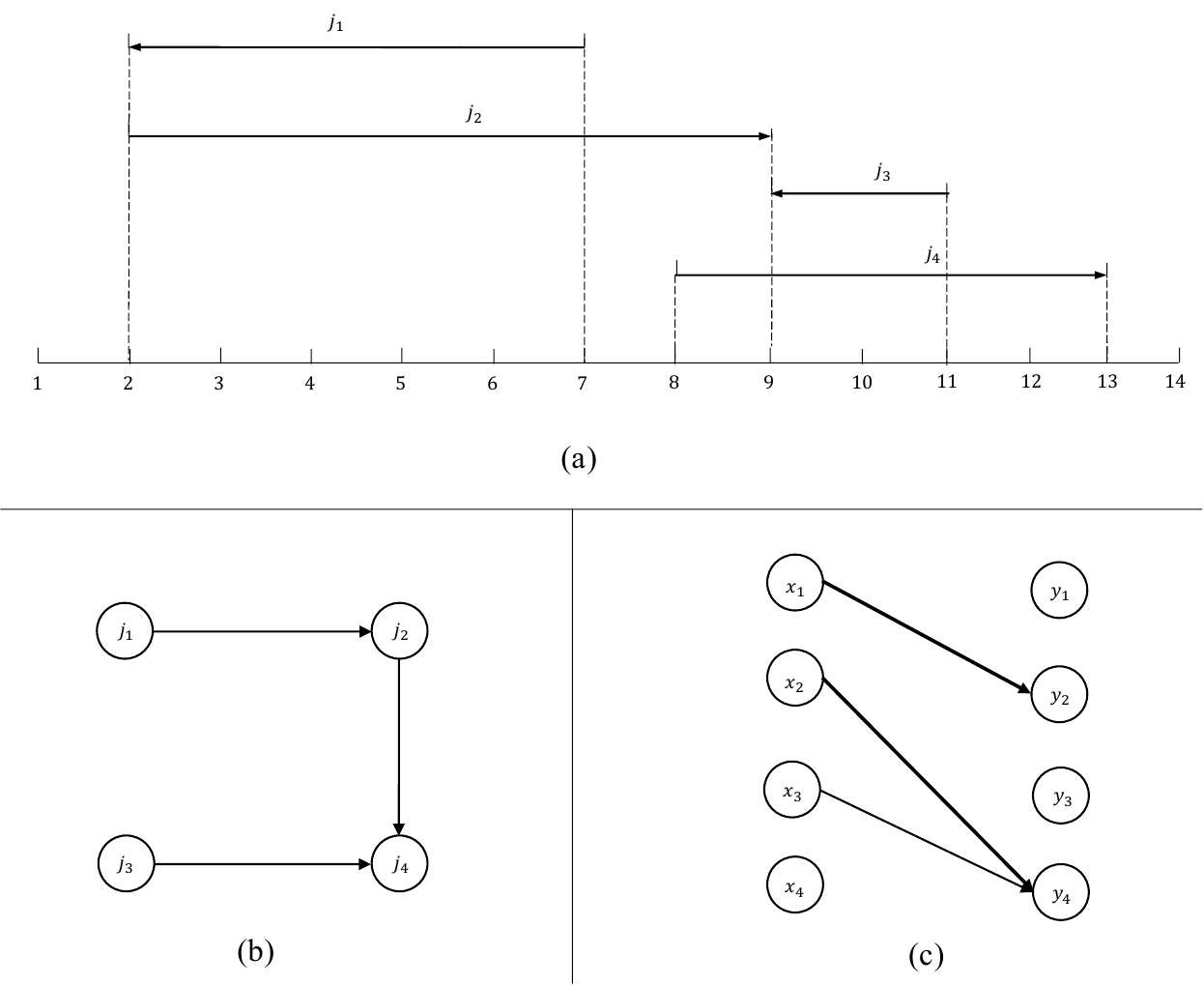}
\caption{Example of arbitrary length case to interval acyclic digraph: (a) input platform; (b) corresponding interval acyclic digraph; (c) corresponding matching problem.}
\label{IntDi2_fig}
\end{figure}

Consider the following example: there are $4$ arbitrary-length jobs with information given in Table \ref{IntDi2_table}. We transform the job information to an interval digraph, as shown in Figure \ref{IntDi2_fig}. Figure \ref{IntDi2_fig}(a) shows the corresponding platform based on the job information in Table \ref{IntDi2_table}.

\begin{table}[htbp]
\centering
\captionsetup{justification=centering}
\caption{Corresponding job information of Figure \ref{IntDi2_fig}}
\setlength{\tabcolsep}{4mm}{
\begin{tabular}{|c|c|c|}
\hline
Job $j$ & Source Set $S_j$ & Terminal Set $T_j$\\ 
\hline\hline
\textbf{$j_1$} & \{7\} & [1,3]  \\ 
\hline
\textbf{$j_2$} & \{2\} & [8,10]  \\ 
\hline
\textbf{$j_3$} & \{11\} & [8,10]  \\ 
\hline
\textbf{$j_4$} & \{8\} & [12,14] \\ 
\hline
\end{tabular}}
\label{IntDi2_table}
\end{table}

Figure \ref{IntDi2_fig} shows how the example in Table \ref{IntDi2_table} is transformed into the interval acyclic digraph. Each job is denoted as a vertex in the interval digraph and each energy-free processing pair is denoted as a directed edge in the interval digraph. Taking $(j_2,j_4)$ as an example, since $t_2=9$, after processing $j_2$, the crane can store energy temporarily in the interval $[t_2-e,t_2+e]$, that is $[8,10]$. If job $j_4$ has not been processed, the origin $s_4=8$, which is exactly in the interval $[8,10]$. This implies that the crane can move one unit-length and then continue processing job $j_4$ without consuming extra energy. The corresponding interval digraph in Figure \ref{IntDi2_fig}(b) is cycle-free, which allows us to reduce the problem to a maximum matching problem by constructing the corresponding bipartite graph in Figure \ref{IntDi2_fig}(c), where the bold line represents the matching edges. The minimum vertex-disjoint path cover requires $\rho =|V|-\mu =2$ paths, and for the scheduling problem, the total energy consumption is also $2$.

\end{document}